\documentclass[twocolumn,showpacs,preprintnumbers,amsmath,amssymb, pr]{revtex4-1}

\usepackage{stmaryrd}
\usepackage{txfonts}
\usepackage{amssymb}
\usepackage{mathrsfs}
\usepackage{graphicx}
\usepackage{dcolumn}
\usepackage{bm}
\usepackage{epsfig}

\begin{document}

\title{Thermally Assisted Penetration and Exclusion of Single Vortex in Mesoscopic Superconductors}

\author{Shi-Zeng Lin\(^{1}\), Takahiro Nishio\(^{2}\), Lev N. Bulaevskii\(^{1}\), Matthias J. Graf\(^{1}\) and Yukio Hasegawa\(^{3}\)}

\affiliation{\(^{1}\)Theoretical Division, Los Alamos National Laboratory, Los Alamos, New Mexico 87545, USA\\
\(^{2}\)Advanced Device Laboratory, RIKEN Advanced Science Institute, 2-1 Hirosawa, Wako,
Saitama 351-0198, Japan\\
\(^{3}\)The Institute for Solid State Physics, The University of Tokyo, Kashiwa-no-ha,
Kashiwa 277-8581, Japan}

\date{\today}

\begin{abstract}
A single vortex overcoming the surface barrier in a mesoscopic superconductor with lateral dimensions of several coherence lengths and thickness of several nanometers provides an ideal platform to study thermal activation of a single vortex. In the presence of thermal fluctuations, there is non-zero probability for vortex penetration into or exclusion from the superconductor even when the surface barrier does not vanish. We consider the thermal activation of a single vortex in a mesoscopic superconducting disk of circular shape. To obtain statistics for the penetration and exclusion magnetic fields, slow and periodic magnetic fields are applied to the superconductor. We calculate the distribution of the penetration and exclusion fields from the thermal activation rate. This distribution can also be measured experimentally, which allows for a quantitative comparison. 
\end{abstract}

 \maketitle

\section{Introduction}
Thermal activation of macroscopic quantum objects over an energy barrier is ubiquitous in condensed matter physics. One well established example is the thermally assisted escape of phase particle in a Josephson junction when the bias current is close to the critical current corresponding to a vanishing barrier\cite{Clarke88}. Combined with analytical calculations and experiments, a quantitative description has been achieved in this case. Magnetic vortices as topological excitations in superconductors are difficult to excite thermally because of their large self-energy compared to that of phase particle in Josephson junctions. For a large of number of vortices the probability of thermal activation increases, which allows for an experimental observation. Thermal activation of bundles of vortices, such as creep motion of vortices\cite{TinkhamBook}, has been observed long time ago. However, no detailed quantitative experimental study of thermal activation of a single vortex has been reported so far because the probability of thermal activation for a single vortex is usually small.

Vortices overcoming the surface barrier in finite-size superconductors has attracted considerable interest, because of the possible observation of thermal activation of a quantum object. For a superconductor with finite size in lateral dimensions, there exists a surface barrier known as the Bean-Livingston (BL) barrier\cite{Bean64}. When the applied magnetic field increases, the BL barrier diminishes and finally disappears at the penetration field $H_p$, where vortex enters the superconductor. The penetration field is larger than $H_{c1}$, the thermodynamic lower critical field for type II superconductors without surface. For the same reason, the exclusion field $H_e$, when vortex leaves the superconductor, is smaller than $H_{c1}$. Thermally assisted penetration of vortices in a thick superconductor was investigated theoretically by Petukhov and Chechetkin decades ago. \cite{Petukhov74} They found that the thermal activation is practically impossible for vortex. For high-$T_c$ superconductors, Kopylov \emph{et. al.} found experimentally that the penetration field decreases when the sample temperature is increased\cite{Kopylov90}. The thermal effect on the penetration and exclusion field in high-$T_c$ superconductors has also been studied in Refs. \cite{Burlachkov93,Burlachkov94,Lewis95}. The importance of thermal fluctuations in these cases can be estimated using the Ginzburg number $G_i^{3D}=\frac{1}{2}({k_B T_c}/[{H_c(0)\xi(0)^3}])^2$, where $k_B$ is the Boltzmann constant, $T_c$ the critical temperature, $H_c(0)$ the thermodynamic critical field and $\xi(0)$ the coherence length at zero temperature\cite{Blatter94}. For conventional superconductors, $G_i^{3D}\approx 10^{-10}$ and thermal fluctuations are weak; while for high-$T_c$ superconductors,  $G_i^{3D}\approx 10^{-2}$ and thermal fluctuations become important. 

For a large size superconductor studied in Refs. \cite{Petukhov74,Kopylov90,Burlachkov93,Burlachkov94,Lewis95}, many vortices penetrate into the superconductor simultaneously once the applied field reaches the penetration field. To observe single vortex penetration, the lateral size of the superconductor should be of several coherence lengths only. In these mesoscopic superconductors, a single vortex enters or exists when the applied magnetic field is tuned. Due to the advances in microfabrication techniques, penetration and exclusion of a single vortex by adjusting applied magnetic field has been observed repeatedly in many experiments\cite{Geim97,Nishio08,Cren09}. In Ref. \cite{Geim97}, the penetration and exclusion of vortex is measured by the Hall probe and in Refs. \cite{Nishio08,Cren09} it is measured by a scanning tunnelling microscope (STM) tip. However, no statistics on the penetration and exclusion field has been carried out to investigate the thermal activation in these studies.

For mesoscopic superconductors with several atomic layers, the reduction of the dimensions promotes the thermal fluctuations. Even for mesoscopic superconductors made of conventional superconductors, $G_i^{2D}=\frac{1}{2}({k_B T_c}/[{H_c(0)\xi(0)^2 d}])^2\approx 10^{-4}$ with the film thickness $d\sim 1\rm{\ nm}$, which is greatly enhanced by a factor $(\xi/d)^2\gg 1$ compared to the bulk value\cite{Pogosov10}. By solving the Ginzburg-Landau equation and the Fokker-Planck equation analytically for a superconducting disk, Pogosov found that the thermal activation of vortex over the BL barrier becomes feasible in mesoscopic superconductors\cite{Pogosov10}. The effects of thermal fluctuations on the penetration of single vortex have also been studied numerically by solving the time-dependent Ginzburg-Landau equations with noise term\cite{Hernandez05}. They observed that the number of vortices in superconductors is fluctuating due to thermal activation when $G_i^{3D}$ is large in simulations.

\begin{figure*}[t]
\psfig{figure=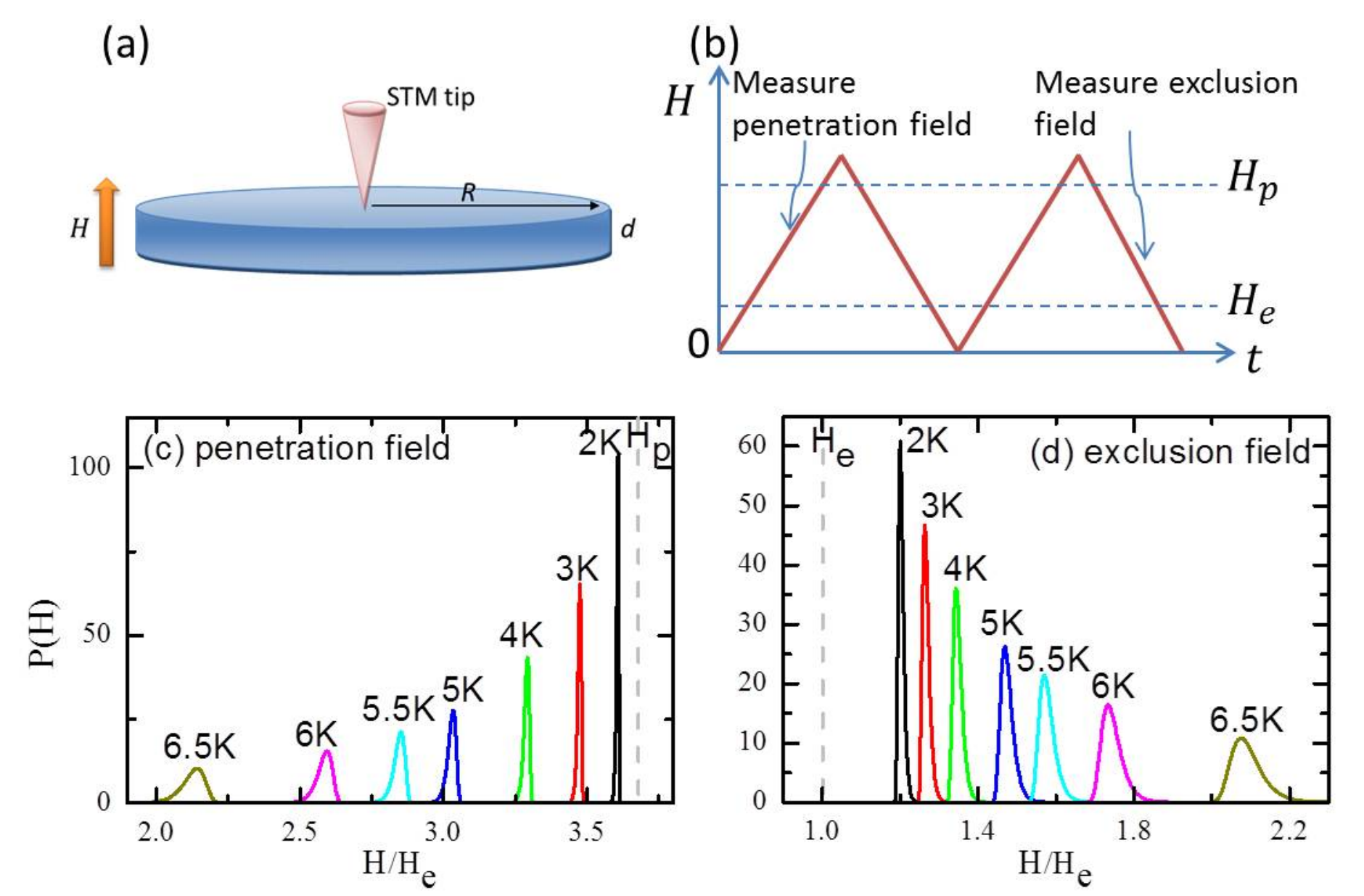,width=17cm} \caption{\label{f1} (color online). (a): A superconducting disk with radius of several coherence lengths and thickness of several nanometers. A STM tip is positioned at the center of the disk to check the presence of a vortex by measuring the zero-bias conductivity. (b): Profiles of the time dependent applied magnetic field. The maximal magnetic field is larger than the mean-field penetration field to initialize a vortex at the center of the disk. In the branch with increasing $H$, one measures the penetration field and in the branch with decreasing field, one measures the exclusion field. (c) and (d): Distribution of the (c) penetration field and (d) exclusion field $P(H)$ at several temperatures. The mean-field penetration and exclusion field is represented by a dashed line. The distribution profiles are obtained using Eqs. (\ref{eq22b}), (\ref{eq30}) and (\ref{eq32}).}
\end{figure*}

In this paper, we consider the thermally assisted penetration and exclusion of a single vortex in a small superconducting disk. We find the distribution of penetration (exclusion) field when the applied magnetic field increases (decreases) adiabatically, as a function of disk size and temperature. The setup and main results are presented in Sec. II. In Sec. III, we derive the BL barrier for a superconducting disk by solving the London equation, while neglecting thermal fluctuations. From the BL barrier, we obtain the mean-field penetration and exclusion field for a single vortex. In Sec. IV, we calculate the thermal activation rate for vortex penetration and exclusion by solving the corresponding Fokker-Planck equation for vortex-as-a-particle and derive the corresponding distribution functions. The paper is closed by a discussion in Sec V.

\section{Proposed measurement}
We consider a circular disk of several coherence lengths in lateral size and several nanometers in thickness $d$ with $d\ll \xi$, where the thermally assisted penetration and exclusion of a single vortex can be observed experimentally, see Fig. \ref{f1}(a). An STM tip is positioned at the center of the disk to check whether the vortex is present or not, by measuring the zero-bias conductivity\cite{Nishio08}. Without thermal fluctuations, the vortex penetrates into the disk when the applied field is larger than the mean-field penetration field. With thermal activation, the probability of vortex penetration is non-zero even when the applied magnetic is smaller than the mean-field penetration field. To find the distribution of penetration field, periodically varying magnetic fields are applied perpendicular to the disk. The setup with a static magnetic field has already been implemented in Refs. \cite{Nishio08,Cren09,Takahiro10}

For the vortex penetration, one changes the applied magnetic field periodically and slowly, see Fig. \ref{f1}(b). To present analytical calculations, we choose the sawtooth wave. One records the magnetic field when vortex enters in the branch with increasing field. By repeating the measurement, one then obtains a distribution of the penetration field at a given temperature. We calculate this distribution function analytically [see Eq. (\ref{eq32}) below], which allows for a quantitative comparison. The measurement of the thermally assisted vortex exclusion is similar. It is determined from the branch with decreasing field, see Fig. \ref{f1}(b).

\section{Energy barrier}

In this section, we derive the mean-field surface barrier for vortex penetration and exclusion in a superconducting disk without accounting for thermal fluctuations in a static magnetic field. The thickness $d$ of the disk is much smaller than the London penetration depth $d\ll \lambda$. For such thin films, the screening of magnetic fields is weak and the effective penetration depth is given by the Pearl length\cite{Pearl64} $\Lambda =2\lambda ^2/d$. The coherence length is also assumed to be much smaller than the penetration depth $\xi\ll\lambda$, and thus we can use the London approximation, where the amplitude of the superconducting order parameter is uniform in space except for the vortex core. The magnetic field distribution $h_z(x,y)$ associated with a vortex is given by the London equation\cite{TinkhamBook}
\begin{equation}\label{eq1}
h_z+\frac{2\pi\Lambda } {c}(\nabla \times \mathbf{g})\cdot\mathbf{\hat{e}_z}=\Phi _0\delta \left(\mathbf{r}-\mathbf{r_v}\right),
\end{equation}
where $\mathbf{g}$ is the sheet current density, $\mathbf{\hat{e}_z}$ a unit vector along the $z$ direction and $\Phi_0=h c/2e$ is the quantum flux. Here $\mathbf{r_v}$ is the position of the vortex and we only consider a single vortex. For the superconducting Pb film in Ref.\cite{Takahiro10} $\Lambda\approx 20 \text{$\ \mu$m}$. The lateral size of the film is much smaller than the Pearl length $R\ll\Lambda$. In this case, the screening of the magnetic field is negligible. We can neglect the first term at the left-hand side of Eq. (\ref{eq1}). For convenience, we introduce a scalar stream function $G$, such that\cite{Brandt95}
\begin{equation}\label{eq2}
\mathbf{g}=\nabla \times \left(G\mathbf{\hat{e}_z}\right).
\end{equation}
At the boundary, the component of $\mathbf{g}$ normal to the boundary is zero, which requires that $G$ is constant at the boundary. Without loss of generality, we take $G=0$ at the edge of the disk. Then Eq. (\ref{eq1}) is reduced to the Poisson equation and the problem is equivalent to the one in electrostatics,
\begin{equation}\label{eq3}
\nabla ^2G=-\frac{c\Phi _0}{2\pi  \Lambda }\delta \left(\mathbf{r}-\mathbf{r_v}\right).
\end{equation}
For a circular disk with radius $R$, we have\cite{Fetter80,Baelus04}
\begin{equation}\label{eq4}
G(\mathbf{r})=\frac{c\Phi _0}{4\pi^2  \Lambda }\ln \left(\left|\frac{\mathbf{r}-\left(R\left/r_v\right.\right)^2\mathbf{r_v}}{\mathbf{r}-\mathbf{r_v}}\right|^2\frac{r_v}{R}\right).
\end{equation}
The energy of the vortex is then given by
\begin{equation}\label{eq5}
\epsilon(\mathbf{r_v})=\frac{\Phi _0}{8\pi }h_z(\mathbf{r}\rightarrow \mathbf{r_v})=\frac{\Phi _0}{2c}G(\mathbf{r}\rightarrow \mathbf{r_v}).
\end{equation}
The energy $\epsilon$ diverges at $\mathbf{r}=\mathbf{r_v}$ and to avoid the divergence, we introduce the standard cutoff $\xi$ near the vortex core, $|\mathbf{r}-\mathbf{r_v}|=\xi$. Then the energy of vortex at $\mathbf{r}=\mathbf{r_v}$ reads
\begin{equation}\label{eq6}
\epsilon(\mathbf{r_v})=\frac{\Phi _0^2}{8\pi ^2\Lambda }\left[\ln \left(1-\left(\frac{r_v}{R}\right)^2\right)+\ln \left(\frac{R}{\xi }\right)\right].
\end{equation}
We need also to add the interaction between the applied field $H$ and the vortex, $U_z=-M_v H$ with the magnetic moment of the vortex\cite{Kogan94,Kogan07,Stejic94}
\begin{equation}\label{eq7}
M_v=\frac{1}{2c}\int dr (\mathbf{r}\times \mathbf{g})\cdot\mathbf{\hat{e}_z}.
\end{equation}
Then we arrive at the total energy of a vortex in a circular disk of radius $R$
\begin{equation}\label{eq8}
U(r_v)=\frac{\Phi _0^2}{8\pi ^2\Lambda }\left[\ln \left(1-\frac{r_v^2}{R^2}\right)+\ln \left(\frac{R}{\xi }\right)-\frac{\pi  R^2 H}{\Phi _0}\left(1-\frac{r_v^2}{R^2}\right)\right].
\end{equation}
Equation (\ref{eq8}) is valid when the vortex is not too close to the edge. For a vortex close to the edge, $R-r_v<\xi$, one should account for the finite size of the vortex core. When $R$ is of the order $\xi$, a different approach, such as direct solution of the Ginzburg-Landau equation, is needed\cite{Pogosov10}. The change of the magnetization when a vortex is at the center of the disk is
 \begin{equation}\label{eq8a}
M_v\equiv-\partial_H U= \frac{\Phi _0 R^2}{8\pi \Lambda }.
\end{equation}
Thus the penetration and exclusion of a single vortex can be determined by measuring $M_v$ using the Hall probe. \cite{Geim97}

A vortex can be trapped in the disk if there is a local minimum at $r_v=0$. Local minimum develops when the magnetic field is larger than the mean-field exclusion field $H_e$,
\begin{equation}\label{eq13b}
H_e=\frac{\Phi_0}{\pi R^2}.
\end{equation}
Above $H_e$ there is a barrier for vortex exclusion. The position of the barrier is determined by the condition when ${\partial U}/{\partial \rho }=0$ with $\rho\equiv r_v/R$, which gives
\begin{equation}\label{eq9}
\rho_b ^2=1-\frac{H_e}{H}.
\end{equation}
It follows that the height of the potential at $\rho_b$ is
\begin{equation}\label{eq10}
U\left(\rho _b\right)=\frac{\Phi _0^2}{8\pi ^2\Lambda }\left[\ln \left(\frac{\Phi _0}{\pi  R^2H }\right)+\ln \left(\frac{R}{\xi }\right)-1\right].
\end{equation}
When $U\left(\rho _b\right)=0$, the barrier for vortex penetration vanishes, because the energy of a vortex at the edge is $U(r_v=R)=0$. This gives the penetration field $H_p$
\begin{equation}\label{eq12}
H_p=\frac{\Phi _0}{\pi  e \xi  R}=\frac{R}{e\xi} H_e,
\end{equation}
with $e=2.718$. At fields above $H_{p}$, nucleation of vortex into the disk becomes favorable. When the radius $R$ increases, the surface barrier decreases thus $H_p$ also decreases. When $U(\rho=0)=0$, a vortex at the center of the disk becomes the ground state, which gives another characteristic field $H_{g}$
\begin{equation}\label{eq13}
H_{g}=\frac{\Phi _0}{\pi R^2}\ln \left(\frac{R}{\xi }\right).
\end{equation}

\begin{figure}[t]
\psfig{figure=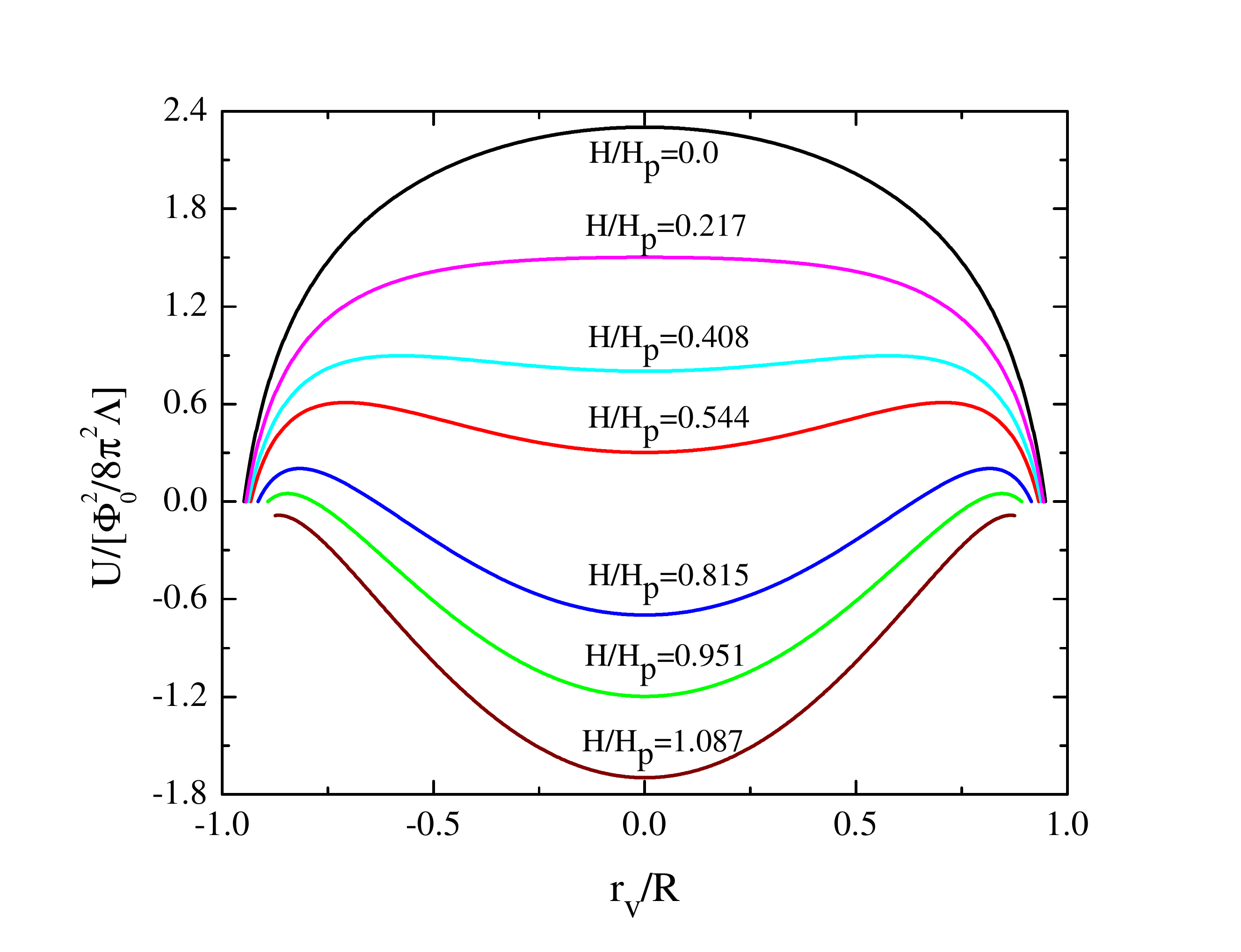,width=\columnwidth} \caption{\label{f2} (color online). Profile of the energy barrier for a vortex in a circular disk with radius $R/\xi=10$ at several typical fields. The penetration field is $H_p=0.074H_{c2}$ and the exclusion field is $H_e=0.02H_{c2}$, with $H_{c2}=\Phi_0/(2\pi\xi^2)$ the thermodynamic upper critical field.}
\end{figure}

For $R=10\xi=300\rm{\ nm}$, we have $H_e\approx 7$ mT and $H_p\approx 26$ mT. Several typical profiles of $U(r_v)$ are displayed in Fig. \ref{f2}. Slightly above $H_e$, a local minimum appears at $r_v=0$, and $U(R)<U(0)$. When the field is increased above $H_{g}$, the local minimum becomes the global minimum and $U(R)>U(0)$. Above $H_p$, the surface barrier for vortex penetration vanishes and the only stable solution is a vortex at the center of the disk.

 \begin{figure}[b]
\psfig{figure=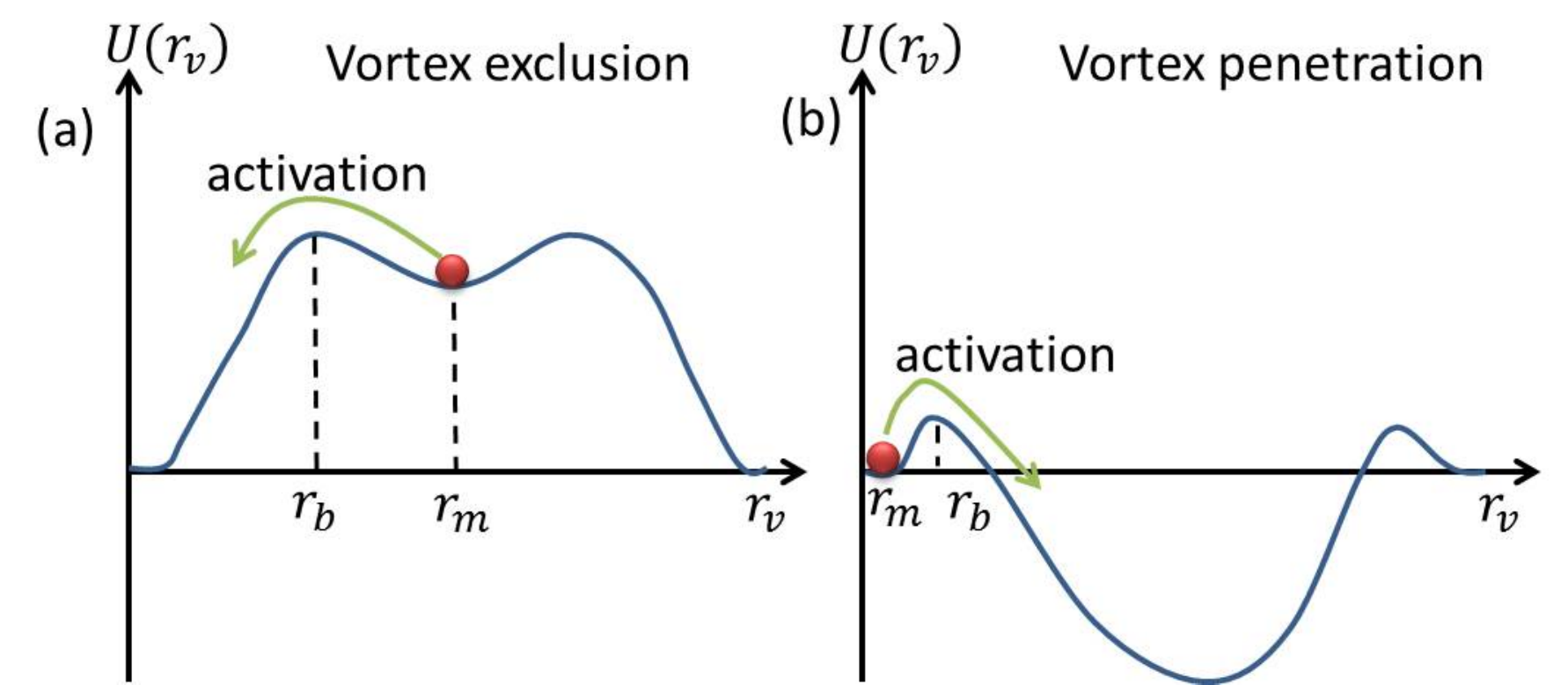,width=\columnwidth} \caption{\label{f3} (color online). Schematic view of thermally assisted vortex exclusion when $H_e<H<H_{g}$ (a) and (b) penetration when $H_{g}<H<H_p$ in a potential $U(r_v)$.}
\end{figure} 

\section{Thermal activation}
When $H$ approaches $H_p$ from below, the barrier decreases and thermal fluctuations will promote the penetration of a vortex. Thus in the presence of fluctuations, a vortex enters at magnetic fields below $H_p$. For vortex exclusion, the vortex leaves the superconductor at fields above the mean-field value $H_e$ due to thermal activation. 

In the presence of time-dependent magnetic field as in Fig. \ref{f1}(b), the energy barrier changes with time in this case. Therefore it is more appropriate to describe the vortex motion in terms of a Lagrangian. The Lagrangian of the vortex is
\begin{equation}\label{eq21a}
\mathcal{L}(r_v, t)=\frac{\Phi _0^2}{8\pi ^2\Lambda }\left[\frac{\pi  R^2 H(t)}{\Phi _0}\left(1-\frac{r_v^2}{R^2}\right)-\ln \left(1-\frac{r_v^2}{R^2}\right)-\ln \left(\frac{R}{\xi }\right)\right].
\end{equation}
The mass of the vortex is small and neglected. The dissipation function is $\mathcal{D}=\eta \dot{r_v}^2/2$, where $\eta=\Phi_0^2/(2\pi  \xi^2c^2\mathcal{R}_n)$ is the Bardeen-Stephen dissipation coefficient with $\mathcal{R}_n$ the resistance of the sample just above $T_c$. $r_v\le R$ is the coordinate of the vortex in the radial direction and $\dot{r_v}$ is the radial vortex velocity. The equation of motion for the vortex in the presence of thermal fluctuations is
\begin{equation}\label{eq21}
\eta  \dot{r_v}=\frac{\partial \mathcal{L}(r_v, t)}{\partial r_v}+F_n,
\end{equation}
where we have introduced a noise force $F_n$ to account for the thermal fluctuations. $F_n$ has a zero mean value $\left\langle F_n(t)\right\rangle=0$ and has a Gaussian correlator
\begin{equation}\label{eq22}
\left\langle F_n(t, \mathbf{r}_v)F_n(t',\mathbf{r}_v')\right\rangle =2\eta  k_B T \delta (t-t')\delta(\mathbf{r}_v-\mathbf{r}_v'),
\end{equation}
where $T$ is the temperature. When the change of the applied field is much slower than the thermalization time of the vortex (see Sec. V), we can use the adiabatic approximation to solve Eq. (\ref{eq21}) and $\mathcal{L}(r_v, t)$ can be replaced by $-U(r_v)$. A schematic view for thermally assisted vortex exclusion and penetration is depicted in Fig. \ref{f3}.

Here we have considered thermally assisted penetration and exclusion of a single vortex. For two vortices, the barrier is twice as large as that for a single vortex. Since the probability for thermal activation is much smaller than that of a single vortex, the simultaneous penetration and exclusion of multiple vortices can be avoided.\cite{Bulaevskii11} 

In Eq. (\ref{eq21}), we have approximated the motion of the vortex in the disk by that of a particle in a one dimensional potential. During the penetration or exclusion of vortex, the motion of single vortex is dictated by the force associated with the surface barrier. The diffusion of vortex along the azimuthal direction due to thermal fluctuations thus can be safely neglected.

\subsection{Thermally assisted exclusion}
Let us first calculate the vortex exclusion rate when $H> H_e$ using the Kramers equation for diffusion over barrier\cite{Hanggi90}
\begin{equation}\label{eq22a}
\Gamma=\frac{\sqrt{-U''(r_m)U''(r_b)}}{2\pi  \eta }\exp \left[-\Delta U/ k_B T\right]
\end{equation}  
where $U''(r_b)$ is the second derivative at the surface barrier and $U''(r_m)$ is the second derivative at the energy minimum where the vortex resides initially. The barrier height is $\Delta U\equiv U(r_b)-U(r_m)$. The Kramers equation is valid when $\Delta U\gg k_B T$.

Using Eq. (\ref{eq22a}), we obtain the rate for the vortex exclusion
\begin{equation}\label{eq22b}
\Gamma_{e} =\frac{\epsilon_c}{\pi  \eta  R^2}\sqrt{2 f \left(f-1\right)^2}\exp \left(-v\left[\ln \left(\frac{1}{e f}\right)+f\right]\right)
\end{equation}  
with $f=H/H_e>1$, $\epsilon_c=\Phi_0^2/(8\pi^2\Lambda)$ and $v=\epsilon_c/(k_B T)$. We estimate $\epsilon_c\approx 2000\rm{\ K}$ for $\Lambda\approx 20\rm{\ \mu m}$. Hence $v\gg 1$ in the whole superconducting state $T<T_c\approx 7.2$K. 

The characteristic frequency is $\omega_c={\epsilon_c}/({\pi  \eta  R^2})\approx 10\rm{\ GHz}$ for $R\approx 300\rm{\ nm}$. The rate decreases when the lateral dimension of the superconductor is reduced at a given $H$. More importantly, because $\epsilon_c$ and $v$ decrease linearly with the film thickness $d$, the thermal activation rate increases gigantically because $v$ appears as an exponent. Note that the spurious decrease of $\Gamma_e$, when $H$ is very close to $H_e$, is an artifact, because Eq. (\ref{eq22b}) becomes invalid when $\Delta U\sim k_B T$. In reality, the rate $\Gamma_e$ increases monotonically when $H$ approaches $H_e$.

\subsection{Thermally assisted penetration} 

For vortex penetration, $U''(r_m)$ is not defined at the edges. The Kramers equation becomes inapplicable, and an explicit solution of the Fokker-Planck equation is needed.

We use the Fokker-Planck equation associated with Eq. (\ref{eq21})
\begin{equation}\label{eq23}
\partial _t\rho(r_v,t) =-\partial _{r_v} J(r_v,t),
\end{equation}
where $\rho(r_v, t)$ is the probability density of finding a vortex at $r_v$ and $J$ is the probability current
\begin{equation}\label{eq24}
J(r_v,t)=-D \exp [-U/(k_B T)]\partial _{r_v}\{\exp [U/(k_B T)]\rho \},
\end{equation}
with the diffusion constant $D=k_B T/\eta$. We use the standard boundary condition for the calculations of the rate\cite{Hanggi90}, by assuming a source of vortices at $r_v=R$ and a sink at $r_v=0$.  The total probability of finding a vortex in the region $R-\xi\le r_v\le R$ is unity because vortex is initially thermalized at the edges,
\begin{equation}\label{eq24aa}
\rho\int_0^{2\pi} d\phi\int_{R-\xi}^{R} r_v dr_v=1,
\end{equation}
which gives $\rho(r_v\approx R,t)=1/(2\pi R\xi)$. $R-\xi$ appears in the lower bound of integral in Eq. (\ref{eq24aa}) because of the uncertainty of the vortex coordinate due to the vortex core of finite size. We are interest in the probability for the vortex penetration, where initially no vortex is located at the center of the disk. To find the thermal activation rate, we can assume a sink at the center of the disk and use the boundary condition $\rho(r_v=0,t)=0$. That means we restart the measurement once the vortex enters into the disk.

Here we consider the regime where $\Delta U\gg k_B T$. In this case, the activation rate for vortex is small, thus $J$ and $\partial_t \rho$ are small\cite{Bulaevskii11b}. Then the probability current is almost independent on $r_v$, i.e., $\partial_{r_v} J=0$. Integrating Eq. (\ref{eq24})  from $r_v=0$ to $r_v=R$ and using the boundary condition, we have
\begin{equation}\label{eq25}
D \rho (r_v\approx R)=J\int_{0}^R \exp [U/(k_B T)] \, dr_v,
\end{equation}
where we have used $U(r_v=R)=0$ because the vortex does not experience the surface barrier at the rim. Thus the thermal activation rate is given by
\begin{equation}\label{eq26}
\Gamma_p\equiv 2\pi R J=\frac{D}{\xi}\left[\int_{0}^{R} \exp [U/(k_B T)] \, dr_v\right]^{-1}.
\end{equation}
The characteristic energy scale of the barrier is $\epsilon_c$. Since $v=\epsilon_c/(k_B T)\gg1$, the integrand drops very rapidly and the dominant contribution is from the region near the barrier. We use the saddle point approximation
\begin{equation}\label{eq27}
\frac{U(r_v)}{\epsilon_c}\approx \mathcal{U}(r_b)+\frac{1}{2}\mathcal{U}''(r_b)(r_v-r_b)^2,
\end{equation}
with $\mathcal{U}''(r_b)<0$. Furthermore we use the following approximation because $v\gg 1$
\begin{equation}\label{eq28}
\exp[v \mathcal{U}''(r_b)(r_v-r_b)^2/2]\approx\sqrt{\frac{2\pi}{-v \mathcal{U}''(r_b)}}\delta(r_v-r_b).
\end{equation}
Then the rate becomes
\begin{equation}\label{eq29}
\Gamma_p=\frac{D}{\xi}\exp[-v \mathcal{U}(r_b)]\sqrt{\frac{-v \mathcal{U}''(r_b)}{2\pi}}.
\end{equation}
Let us compare Eq. (\ref{eq29}) to the Kramers formula Eq. (\ref{eq22a}). These two expressions coincide if $U''(r_m)=4\pi k_B T/\xi^2$. For the vortex penetration, the vortex is thermalized at $r_v=R$ and gains kinetic energy $k_B T$. Because of the uncertainty of the vortex position of the order $\xi$, one may assign an oscillation frequency of the order $k_B T/\xi^2$. We then arrive at a consistent description based on the direct calculations of the Fokker-Planck equation and the Kramers equation.

Finally, for the circular disk, we have the rate for vortex penetration
\begin{equation}\label{eq30}
\Gamma_{p}=\frac{D}{R\xi}\left(\frac{\xi e f}{R }\right)^{v}\sqrt{\frac{2 v f \left(f-1\right)}{\pi } },
\end{equation}
with characteristic frequency $\omega_c={D}/{R\xi}\approx 10 \rm{\ GHz}$. Note that both $\Gamma_e$ and $\Gamma_p$ are strongly nonlinear functions of the radius $R$ and the applied field $H$.

\subsection{Results}

\begin{figure}[t]
\psfig{figure=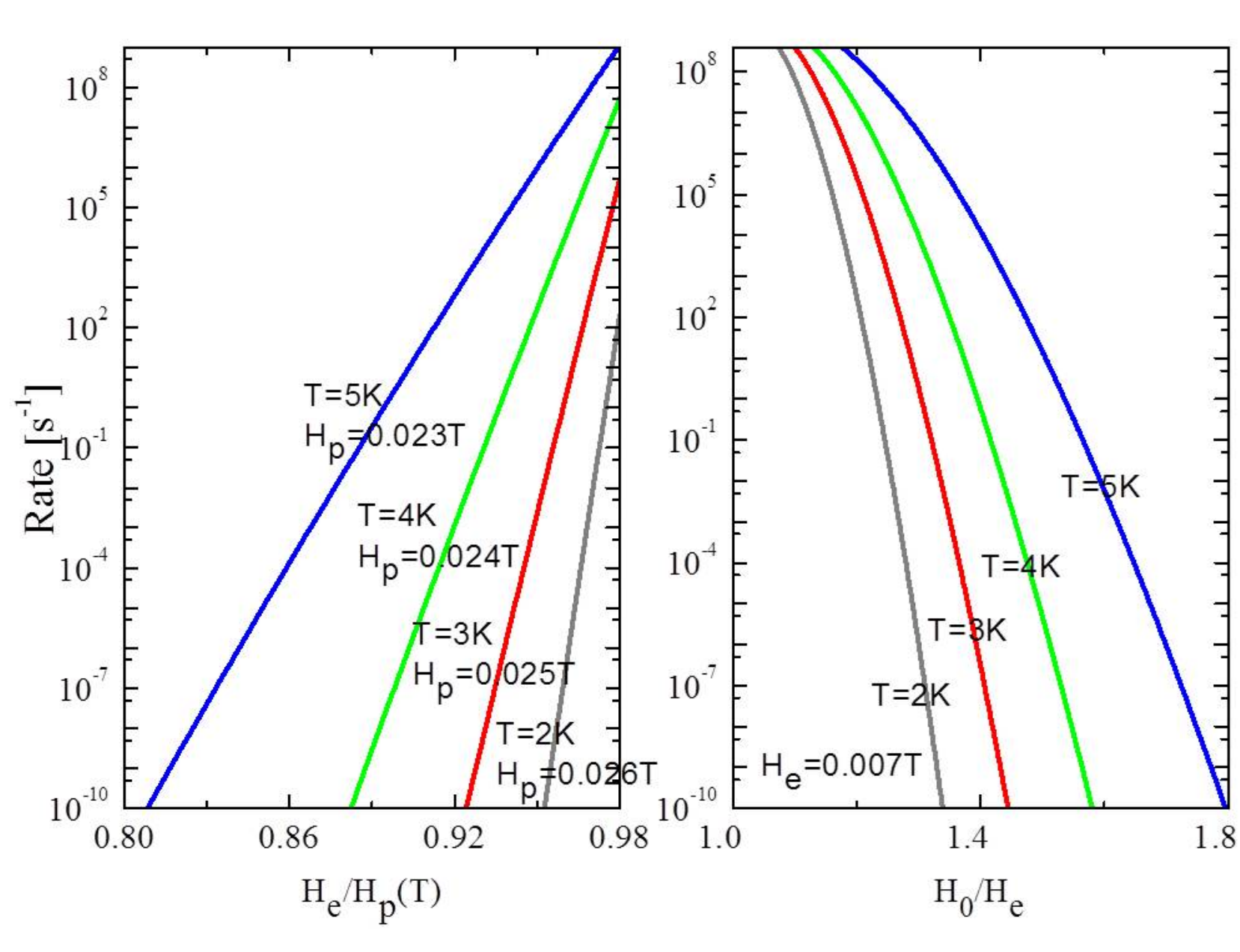,width=\columnwidth} \caption{\label{f5a} (color online). Thermal activation rate for (a) the penetration of a vortex, (b) the exclusion of a vortex. The mean-field penetration and exclusion fields are also shown in the labels. Please note that the mean-field penetration field depends on temperature while the mean-field exclusion field is temperature independent.}
\end{figure}

To estimate the activation rate, we use the parameters from experiments \cite{Takahiro10}. $\lambda=150\rm{\ nm}$, $\xi=30\rm{\ nm}$ at $2$ K and $T_c=7.2$ K. We take $R=300\rm{\ nm}$, $d=2.5\rm{\ nm}$ and $\mathcal{R}_n=400\rm{\ \Omega}$. The temperature dependences of $\xi(T)$ and $\Lambda(T)$ are calculated with the BCS theory. The rate for the vortex penetration and exclusion is shown in Fig. \ref{f5a}.

Experimentally, one measures the distribution of the penetration or exclusion fields when the magnetic field is swept periodically. The distribution of the penetration field can be obtained from the rate in the following way \cite{Fulton74}. The probability for a vortex entering the disk after a waiting time $t$ for the branch with increasing magnetic field in Fig. \ref{f1}(b) is
\begin{equation}\label{eq31}
W[H(t)]=1-\exp \left[-\int_0^t \Gamma_p [H(t')] \, dt'\right].
\end{equation}
Hence the probability distribution of the penetration field is
\begin{equation}\label{eq32}
P(H)=\frac{dW}{d H}=\frac{\Gamma_p [H]}{\dot{H}}\exp \left(-y \right),
\end{equation}
with
\begin{equation}\label{eq32a}
y(H)=\int_0^H \frac{\Gamma_p [H']}{\dot{H}'} \, dH',
\end{equation}
where $\dot{H}$ is the derivative of $H$ with respect to time. For the penetration field measurement, $\dot{H}>0$. In a similar way, one obtains the distribution of the exclusion field for the branch with decreasing magnetic field in Fig. \ref{f1}(b)
\begin{equation}\label{eq33}
P(H)=\frac{\Gamma_e [H]}{|\dot{H}|}\exp \left[-\int_{H}^{+\infty} \frac{\Gamma_e [H']}{|\dot{H}'|} \, dH'\right].
\end{equation}
For the exclusion field measurement, $\dot{H}<0$. Experimentally, the upper limit of the integral is replaced by the maximal magnetic field used in experiment. The difference is negligible because $\Gamma_e\sim \exp(-v H/H_e)\approx 0$ when $H\gg H_e$.

The distributions of the penetration field and exclusion field as a function of temperature are shown in Fig. \ref{f1}(e) and (f). In the calculations, we use $\dot{H}=14\rm{\ mT/s}$ for penetration field and $\dot{H}=-14\rm{\ mT/s}$ for the exclusion field. As the temperature increases, the distribution profile becomes broad. The most probable penetration field decreases with temperature, and deviates from the mean-field value. As $T$ increases, the exclusion field increases and also deviates strongly from the mean-field value. Therefore the hysteretic region between the vortex penetration and exclusion is reduced with increasing temperature.

\section{Discussions}

Let us discuss the effect of time dependent applied field. First, the time dependent magnetic field induces an electric field and current, which generates heat. Secondly, the induced current exerts a Lorentz force on the vortex, thus it gives additional contribution to the surface potential. Thirdly, one should check the validity of the adiabatic approximation used in the evaluation of the rate in Eqs. (\ref{eq22b}) and (\ref{eq30}).

The induced electric field according to the Faraday's Law is $E(r)=-\dot{H} r/(2c)$, and the generated Joule heating is $Q= \sigma_q  \pi \dot{H}^2R^4d/(8c^2)$, where $\sigma_q$ is the conductivity of quasiparticles and is much smaller than the conductivity of the normal state. Using the heat balance equation, 
\begin{equation}
d C \rho_d \partial _tT=\frac{Q}{S}+\alpha \left(T_B-T\right)
\end{equation}
the temperature increase after the system reaches a stationary state is $\Delta T=\sigma_q {\dot{H}^2}R^2d/{(8c^2\alpha )}$, where $\alpha$ is the heat exchange rate with substrate. Here $C$ is the heat capacity and $\rho_d$ is the mass density of the superconductor. One can neglect heating if $\Delta T/T\ll 1$. For the sawtooth wave in Fig. \ref{f1}(b) and (c), $\dot{H}$ is of order of $\omega_H H_p$ with $\omega_H$ being the period of the time-dependent field. To give an order of magnitude estimate, we take $\alpha \approx 10^7\rm{\ W/m^2\cdot K} $ and take the conductivity of Pb film in the normal state above $T_c$, $\sigma_q d=0.01 \rm{\ \Omega^{-1}}$. For $\omega_H\ll 1\rm{\ THz}$, the heating effect is negligibly small.  

The induced current of Cooper pairs by the time-dependent magnetic field is accounted for by the last term on the right-hand side of Eq. (\ref{eq8}). The induced quasiparticles $J_q(r)=-\sigma_q \dot{H} r/(2c)$ tilts additionally the potential for vortex penetration or exclusion. The change of the potential is
\begin{equation}
\Delta U_q=d\int _0^RJ_q \frac{\Phi _0}{c}dr=\frac{d{\sigma_q }\dot{H} {\Phi _0}{R^2}}{4c^2}.
\end{equation}
It is required that $\Delta U_q/\epsilon_c\ll 1$. This condition is satisfied when $\omega_H\ll 1\rm{\ THz}$.

To validate the adiabatic approximation employed in Eqs. (\ref{eq22b}) and (\ref{eq30}), the change of the applied field should be much slower than the thermalization time  of vortex (or the relaxation time of a vortex to reach the energy minimum), such that the vortex remains in equilibrium when the magnetic field changes\cite{Smelyanskiy99}. From Eq. (\ref{eq21}), the thermalization time is $\eta/\partial_x^2 U\approx 1/\omega_c\approx10^{-10}\rm{\ s}$ for the parameters used in the previous section. The adiabatic approximation thus is justified when $\omega_H\ll 10\rm{\ GHz}$. 

Let us discuss the detailed form of the applied magnetic field. The rate description is valid when $\Delta U/(k_B T)\gg 1$. The barrier near $H_p$ can be written as $\Delta U=\epsilon_c(1-H/H_p)$. This gives an upper bound for the applied field $1-H/H_p\gg k_B T/\epsilon_c\approx 1/200$ at $T_c$. This means that the rate description is valid almost in the whole region of the magnetic field below $H_p$. On the other hand, the period of the applied field should be large enough to count the event of vortex penetration in one period, which gives an upper limit for $\omega_H$. The upper limit can be estimated as follows. The probability to observe vortex penetration according to Eq. (\ref{eq31}) is, $W=1-\exp(-y)$. The rate can be written as $\Gamma_p\approx\omega_c \exp(-v (1-H/H_p)]$. Then we have
\begin{equation}
y=-\frac{\omega_c H_p}{v \dot{H}'} \exp\left[-v \left(1-\frac{H}{H_p}\right)\right]\approx\frac{\omega_c}{v \omega_H}\exp(-10),
\end{equation}
where we have taken $v (1-H/H_p)=10$, such that the rate description is valid. In order to have large probability for vortex penetration in one period, $y> 1$, which gives $\omega_H< 10\rm{\ kHz}$. For such frequencies, the heating effect and tilt of the surface barrier are negligible.

Some clues for the thermally assisted penetration and exclusion of a vortex in mesoscopic superconductors can be found by comparing two experiments\cite{Nishio08,Cren09}. In a recent experiment by Cren \emph{et. al.}\cite{Cren09}, they measured the penetration and exclusion of the vortex in Pb films using an STM. No hysteresis in the penetration and exclusion of the vortex was observed. While in a similar experiment by Nishio \emph{et. al.}\cite{Nishio08}, hysteresis was clearly observed. These two experiments use Pb superconducting disks with similar thickness and radius. $R\approx 2\xi\approx 60\rm{\ nm}$ in both experiments and $d=2.5\rm{\ nm}$ in Ref. \cite{Nishio08} and $d=5.5\rm{\ nm}$ in Ref. \cite{Cren09}. The temperature in Ref. \cite{Cren09} is 4.3K while in Ref.\cite{Nishio08} is 2K with $T_c\approx7.2\rm{\ K}$. According to Fig. \ref{f1} (e) and (f), increasing the temperature reduces the hysteresis, which disappears eventually at higher temperature. The thermally assisted penetration and exclusion of vortex thus can qualitatively account for the difference in Ref. \cite{Cren09} and Ref. \cite{Nishio08}.

In summary, we have studied the thermally activated penetration and exclusion of a single vortex in a mesoscopic superconductor. We derived the energy barrier for a vortex as a particle using the London approximation, from which we obtained the mean-field penetration, exclusion field and thermodynamic critical field. We then calculated the thermal activation rate for the vortex penetration and exclusion based on the Fokker-Planck equation and Kramers' escape rate in the adiabatic region of field change. Based on the activation rate, we obtained the distribution of the penetration and exclusion field due to thermal activation. Finally we proposed measurements of the distribution of penetration and exclusion magnetic fields to test our model of thermally assisted single-vortex motion in mesoscopic superconductors.

\section{Acknowledgement}
Work at the Los Alamos National Laboratory was performed under the auspices of the U.S. DOE contract No. DE-AC52-06NA25396 through the LDRD program. Work at RIKEN (TN) was supported by JSPS Grant-in-Aid for Young Scientists (B) No.22760018 and Special Postdoctoral Researchers Program of RIKEN. Work at University of Tokyo (YH) was supported by Grand-in Aid for Scientific Research (21360018), Ministry of Education, Culture, Sports, Science and Technology (MEXT), Japan.


\begin{thebibliography}{27}%
\makeatletter
\providecommand \@ifxundefined [1]{%
 \@ifx{#1\undefined}
}%
\providecommand \@ifnum [1]{%
 \ifnum #1\expandafter \@firstoftwo
 \else \expandafter \@secondoftwo
 \fi
}%
\providecommand \@ifx [1]{%
 \ifx #1\expandafter \@firstoftwo
 \else \expandafter \@secondoftwo
 \fi
}%
\providecommand \natexlab [1]{#1}%
\providecommand \enquote  [1]{``#1''}%
\providecommand \bibnamefont  [1]{#1}%
\providecommand \bibfnamefont [1]{#1}%
\providecommand \citenamefont [1]{#1}%
\providecommand \href@noop [0]{\@secondoftwo}%
\providecommand \href [0]{\begingroup \@sanitize@url \@href}%
\providecommand \@href[1]{\@@startlink{#1}\@@href}%
\providecommand \@@href[1]{\endgroup#1\@@endlink}%
\providecommand \@sanitize@url [0]{\catcode `\\12\catcode `\$12\catcode
  `\&12\catcode `\#12\catcode `\^12\catcode `\_12\catcode `\%12\relax}%
\providecommand \@@startlink[1]{}%
\providecommand \@@endlink[0]{}%
\providecommand \url  [0]{\begingroup\@sanitize@url \@url }%
\providecommand \@url [1]{\endgroup\@href {#1}{\urlprefix }}%
\providecommand \urlprefix  [0]{URL }%
\providecommand \Eprint [0]{\href }%
\providecommand \doibase [0]{http://dx.doi.org/}%
\providecommand \selectlanguage [0]{\@gobble}%
\providecommand \bibinfo  [0]{\@secondoftwo}%
\providecommand \bibfield  [0]{\@secondoftwo}%
\providecommand \translation [1]{[#1]}%
\providecommand \BibitemOpen [0]{}%
\providecommand \bibitemStop [0]{}%
\providecommand \bibitemNoStop [0]{.\EOS\space}%
\providecommand \EOS [0]{\spacefactor3000\relax}%
\providecommand \BibitemShut  [1]{\csname bibitem#1\endcsname}%
\let\auto@bib@innerbib\@empty
\bibitem [{\citenamefont {Clarke}\ \emph {et~al.}(1988)\citenamefont {Clarke},
  \citenamefont {Cleland}, \citenamefont {Devoret}, \citenamefont {Esteve},\
  and\ \citenamefont {Martinis}}]{Clarke88}%
  \BibitemOpen
  \bibfield  {author} {\bibinfo {author} {\bibfnamefont {J.}~\bibnamefont
  {Clarke}}, \bibinfo {author} {\bibfnamefont {A.~N.}\ \bibnamefont {Cleland}},
  \bibinfo {author} {\bibfnamefont {M.~H.}\ \bibnamefont {Devoret}}, \bibinfo
  {author} {\bibfnamefont {D.}~\bibnamefont {Esteve}}, \ and\ \bibinfo {author}
  {\bibfnamefont {J.~M.}\ \bibnamefont {Martinis}},\ }\href@noop {} {\bibfield
  {journal} {\bibinfo  {journal} {Science}\ }\textbf {\bibinfo {volume}
  {239}},\ \bibinfo {pages} {992} (\bibinfo {year} {1988})}\BibitemShut
  {NoStop}%
\bibitem [{\citenamefont {Tinkham}(1996)}]{TinkhamBook}%
  \BibitemOpen
  \bibfield  {author} {\bibinfo {author} {\bibfnamefont {M.}~\bibnamefont
  {Tinkham}},\ }\href@noop {} {\emph {\bibinfo {title} {Introduction to
  Superconductivity}}}\ (\bibinfo  {publisher} {McGraw-Hill, Inc.},\ \bibinfo
  {address} {New York},\ \bibinfo {year} {1996})\BibitemShut {NoStop}%
\bibitem [{\citenamefont {Bean}\ and\ \citenamefont
  {Livingston}(1964)}]{Bean64}%
  \BibitemOpen
  \bibfield  {author} {\bibinfo {author} {\bibfnamefont {C.~P.}\ \bibnamefont
  {Bean}}\ and\ \bibinfo {author} {\bibfnamefont {J.~D.}\ \bibnamefont
  {Livingston}},\ }\href@noop {} {\bibfield  {journal} {\bibinfo  {journal}
  {Physical Review Letters}\ }\textbf {\bibinfo {volume} {12}},\ \bibinfo
  {pages} {14} (\bibinfo {year} {1964})}\BibitemShut {NoStop}%
\bibitem [{\citenamefont {Petukhov}\ and\ \citenamefont
  {Chechetkin}(1974)}]{Petukhov74}%
  \BibitemOpen
  \bibfield  {author} {\bibinfo {author} {\bibfnamefont {B.}~\bibnamefont
  {Petukhov}}\ and\ \bibinfo {author} {\bibfnamefont {V.~R.}\ \bibnamefont
  {Chechetkin}},\ }\href@noop {} {\bibfield  {journal} {\bibinfo  {journal}
  {Sov. Phys. JETP}\ }\textbf {\bibinfo {volume} {38}},\ \bibinfo {pages} {827}
  (\bibinfo {year} {1974})}\BibitemShut {NoStop}%
\bibitem [{\citenamefont {Kopylov}\ \emph {et~al.}(1990)\citenamefont
  {Kopylov}, \citenamefont {Koshelev}, \citenamefont {Schegolev},\ and\
  \citenamefont {Togonidze}}]{Kopylov90}%
  \BibitemOpen
  \bibfield  {author} {\bibinfo {author} {\bibfnamefont {V.~N.}\ \bibnamefont
  {Kopylov}}, \bibinfo {author} {\bibfnamefont {A.~E.}\ \bibnamefont
  {Koshelev}}, \bibinfo {author} {\bibfnamefont {I.~F.}\ \bibnamefont
  {Schegolev}}, \ and\ \bibinfo {author} {\bibfnamefont {T.~G.}\ \bibnamefont
  {Togonidze}},\ }\href@noop {} {\bibfield  {journal} {\bibinfo  {journal}
  {Physica C}\ }\textbf {\bibinfo {volume} {170}},\ \bibinfo {pages} {291}
  (\bibinfo {year} {1990})}\BibitemShut {NoStop}%
\bibitem [{\citenamefont {Burlachkov}(1993)}]{Burlachkov93}%
  \BibitemOpen
  \bibfield  {author} {\bibinfo {author} {\bibfnamefont {L.}~\bibnamefont
  {Burlachkov}},\ }\href@noop {} {\bibfield  {journal} {\bibinfo  {journal}
  {Phys. Rev. B}\ }\textbf {\bibinfo {volume} {47}},\ \bibinfo {pages} {8056}
  (\bibinfo {year} {1993})}\BibitemShut {NoStop}%
\bibitem [{\citenamefont {Burlachkov}\ \emph {et~al.}(1994)\citenamefont
  {Burlachkov}, \citenamefont {Geshkenbein}, \citenamefont {Koshelev},
  \citenamefont {Larkin},\ and\ \citenamefont {Vinokur}}]{Burlachkov94}%
  \BibitemOpen
  \bibfield  {author} {\bibinfo {author} {\bibfnamefont {L.}~\bibnamefont
  {Burlachkov}}, \bibinfo {author} {\bibfnamefont {V.~B.}\ \bibnamefont
  {Geshkenbein}}, \bibinfo {author} {\bibfnamefont {A.~E.}\ \bibnamefont
  {Koshelev}}, \bibinfo {author} {\bibfnamefont {A.~I.}\ \bibnamefont
  {Larkin}}, \ and\ \bibinfo {author} {\bibfnamefont {V.~M.}\ \bibnamefont
  {Vinokur}},\ }\href@noop {} {\bibfield  {journal} {\bibinfo  {journal} {Phys.
  Rev. B}\ }\textbf {\bibinfo {volume} {50}},\ \bibinfo {pages} {16770}
  (\bibinfo {year} {1994})}\BibitemShut {NoStop}%
\bibitem [{\citenamefont {Lewis}\ \emph {et~al.}(1995)\citenamefont {Lewis},
  \citenamefont {Vinokur}, \citenamefont {Wagner},\ and\ \citenamefont
  {Hinks}}]{Lewis95}%
  \BibitemOpen
  \bibfield  {author} {\bibinfo {author} {\bibfnamefont {J.~A.}\ \bibnamefont
  {Lewis}}, \bibinfo {author} {\bibfnamefont {V.~M.}\ \bibnamefont {Vinokur}},
  \bibinfo {author} {\bibfnamefont {J.}~\bibnamefont {Wagner}}, \ and\ \bibinfo
  {author} {\bibfnamefont {D.}~\bibnamefont {Hinks}},\ }\href@noop {}
  {\bibfield  {journal} {\bibinfo  {journal} {Phys. Rev. B}\ }\textbf {\bibinfo
  {volume} {52}},\ \bibinfo {pages} {R3852} (\bibinfo {year}
  {1995})}\BibitemShut {NoStop}%
\bibitem [{\citenamefont {Blatter}\ \emph {et~al.}(1994)\citenamefont
  {Blatter}, \citenamefont {Feigelman}, \citenamefont {Geshkenbein},
  \citenamefont {Larkin},\ and\ \citenamefont {Vinokur}}]{Blatter94}%
  \BibitemOpen
  \bibfield  {author} {\bibinfo {author} {\bibfnamefont {G.}~\bibnamefont
  {Blatter}}, \bibinfo {author} {\bibfnamefont {M.~V.}\ \bibnamefont
  {Feigelman}}, \bibinfo {author} {\bibfnamefont {V.~B.}\ \bibnamefont
  {Geshkenbein}}, \bibinfo {author} {\bibfnamefont {A.~I.}\ \bibnamefont
  {Larkin}}, \ and\ \bibinfo {author} {\bibfnamefont {V.~M.}\ \bibnamefont
  {Vinokur}},\ }\href@noop {} {\bibfield  {journal} {\bibinfo  {journal} {Rev.
  Mod. Phys.}\ }\textbf {\bibinfo {volume} {66}},\ \bibinfo {pages} {1125}
  (\bibinfo {year} {1994})}\BibitemShut {NoStop}%
\bibitem [{\citenamefont {Geim}\ \emph {et~al.}(1997)\citenamefont {Geim},
  \citenamefont {Grigorieva}, \citenamefont {Dubonos}, \citenamefont {Lok},
  \citenamefont {Maan}, \citenamefont {Filippov},\ and\ \citenamefont
  {Peeters}}]{Geim97}%
  \BibitemOpen
  \bibfield  {author} {\bibinfo {author} {\bibfnamefont {A.~K.}\ \bibnamefont
  {Geim}}, \bibinfo {author} {\bibfnamefont {I.~V.}\ \bibnamefont
  {Grigorieva}}, \bibinfo {author} {\bibfnamefont {S.~V.}\ \bibnamefont
  {Dubonos}}, \bibinfo {author} {\bibfnamefont {J.~G.~S.}\ \bibnamefont {Lok}},
  \bibinfo {author} {\bibfnamefont {J.~C.}\ \bibnamefont {Maan}}, \bibinfo
  {author} {\bibfnamefont {A.~E.}\ \bibnamefont {Filippov}}, \ and\ \bibinfo
  {author} {\bibfnamefont {F.~M.}\ \bibnamefont {Peeters}},\ }\href@noop {}
  {\bibfield  {journal} {\bibinfo  {journal} {Nature}\ }\textbf {\bibinfo
  {volume} {390}},\ \bibinfo {pages} {259} (\bibinfo {year}
  {1997})}\BibitemShut {NoStop}%
\bibitem [{\citenamefont {Nishio}\ \emph {et~al.}(2008)\citenamefont {Nishio},
  \citenamefont {An}, \citenamefont {Nomura}, \citenamefont {Miyachi},
  \citenamefont {Eguchi}, \citenamefont {Sakata}, \citenamefont {Lin},
  \citenamefont {Hayashi}, \citenamefont {Nakai}, \citenamefont {Machida},\
  and\ \citenamefont {Hasegawa}}]{Nishio08}%
  \BibitemOpen
  \bibfield  {author} {\bibinfo {author} {\bibfnamefont {T.}~\bibnamefont
  {Nishio}}, \bibinfo {author} {\bibfnamefont {T.}~\bibnamefont {An}}, \bibinfo
  {author} {\bibfnamefont {A.}~\bibnamefont {Nomura}}, \bibinfo {author}
  {\bibfnamefont {K.}~\bibnamefont {Miyachi}}, \bibinfo {author} {\bibfnamefont
  {T.}~\bibnamefont {Eguchi}}, \bibinfo {author} {\bibfnamefont
  {H.}~\bibnamefont {Sakata}}, \bibinfo {author} {\bibfnamefont {S.~Z.}\
  \bibnamefont {Lin}}, \bibinfo {author} {\bibfnamefont {N.}~\bibnamefont
  {Hayashi}}, \bibinfo {author} {\bibfnamefont {N.}~\bibnamefont {Nakai}},
  \bibinfo {author} {\bibfnamefont {M.}~\bibnamefont {Machida}}, \ and\
  \bibinfo {author} {\bibfnamefont {Y.}~\bibnamefont {Hasegawa}},\ }\href@noop
  {} {\bibfield  {journal} {\bibinfo  {journal} {Phys. Rev. Lett.}\ }\textbf
  {\bibinfo {volume} {101}},\ \bibinfo {pages} {167001} (\bibinfo {year}
  {2008})}\BibitemShut {NoStop}%
\bibitem [{\citenamefont {Cren}\ \emph {et~al.}(2009)\citenamefont {Cren},
  \citenamefont {Fokin}, \citenamefont {Debontridder}, \citenamefont {Dubost},\
  and\ \citenamefont {Roditchev}}]{Cren09}%
  \BibitemOpen
  \bibfield  {author} {\bibinfo {author} {\bibfnamefont {T.}~\bibnamefont
  {Cren}}, \bibinfo {author} {\bibfnamefont {D.}~\bibnamefont {Fokin}},
  \bibinfo {author} {\bibfnamefont {F.}~\bibnamefont {Debontridder}}, \bibinfo
  {author} {\bibfnamefont {V.}~\bibnamefont {Dubost}}, \ and\ \bibinfo {author}
  {\bibfnamefont {D.}~\bibnamefont {Roditchev}},\ }\href@noop {} {\bibfield
  {journal} {\bibinfo  {journal} {Phys. Rev. Lett.}\ }\textbf {\bibinfo
  {volume} {102}},\ \bibinfo {pages} {127005} (\bibinfo {year}
  {2009})}\BibitemShut {NoStop}%
\bibitem [{\citenamefont {Pogosov}(2010)}]{Pogosov10}%
  \BibitemOpen
  \bibfield  {author} {\bibinfo {author} {\bibfnamefont {W.~V.}\ \bibnamefont
  {Pogosov}},\ }\href@noop {} {\bibfield  {journal} {\bibinfo  {journal} {Phys.
  Rev. B}\ }\textbf {\bibinfo {volume} {81}},\ \bibinfo {pages} {184517}
  (\bibinfo {year} {2010})}\BibitemShut {NoStop}%
\bibitem [{\citenamefont {Hernandez}\ \emph {et~al.}(2005)\citenamefont
  {Hernandez}, \citenamefont {Baelus}, \citenamefont {Dominguez},\ and\
  \citenamefont {Peeters}}]{Hernandez05}%
  \BibitemOpen
  \bibfield  {author} {\bibinfo {author} {\bibfnamefont {A.~D.}\ \bibnamefont
  {Hernandez}}, \bibinfo {author} {\bibfnamefont {B.~J.}\ \bibnamefont
  {Baelus}}, \bibinfo {author} {\bibfnamefont {D.}~\bibnamefont {Dominguez}}, \
  and\ \bibinfo {author} {\bibfnamefont {F.~M.}\ \bibnamefont {Peeters}},\
  }\href@noop {} {\bibfield  {journal} {\bibinfo  {journal} {Phys. Rev. B}\
  }\textbf {\bibinfo {volume} {71}},\ \bibinfo {pages} {214524} (\bibinfo
  {year} {2005})}\BibitemShut {NoStop}%
\bibitem [{\citenamefont {Nishio}\ \emph {et~al.}(2010)\citenamefont {Nishio},
  \citenamefont {Lin}, \citenamefont {An}, \citenamefont {Eguchi},\ and\
  \citenamefont {Hasegawa}}]{Takahiro10}%
  \BibitemOpen
  \bibfield  {author} {\bibinfo {author} {\bibfnamefont {T.}~\bibnamefont
  {Nishio}}, \bibinfo {author} {\bibfnamefont {S.~Z.}\ \bibnamefont {Lin}},
  \bibinfo {author} {\bibfnamefont {T.}~\bibnamefont {An}}, \bibinfo {author}
  {\bibfnamefont {T.}~\bibnamefont {Eguchi}}, \ and\ \bibinfo {author}
  {\bibfnamefont {Y.}~\bibnamefont {Hasegawa}},\ }\href@noop {} {\bibfield
  {journal} {\bibinfo  {journal} {Nanotechnology}\ }\textbf {\bibinfo {volume}
  {21}},\ \bibinfo {pages} {465704} (\bibinfo {year} {2010})}\BibitemShut
  {NoStop}%
\bibitem [{\citenamefont {Pearl}(1964)}]{Pearl64}%
  \BibitemOpen
  \bibfield  {author} {\bibinfo {author} {\bibfnamefont {J.}~\bibnamefont
  {Pearl}},\ }\href@noop {} {\bibfield  {journal} {\bibinfo  {journal} {Appl.
  Phys. Lett.}\ }\textbf {\bibinfo {volume} {5}},\ \bibinfo {pages} {65}
  (\bibinfo {year} {1964})}\BibitemShut {NoStop}%
\bibitem [{\citenamefont {Brandt}(1995)}]{Brandt95}%
  \BibitemOpen
  \bibfield  {author} {\bibinfo {author} {\bibfnamefont {E.~H.}\ \bibnamefont
  {Brandt}},\ }\href@noop {} {\bibfield  {journal} {\bibinfo  {journal} {Phys.
  Rev. Lett.}\ }\textbf {\bibinfo {volume} {74}},\ \bibinfo {pages} {3025}
  (\bibinfo {year} {1995})}\BibitemShut {NoStop}%
\bibitem [{\citenamefont {Fetter}(1980)}]{Fetter80}%
  \BibitemOpen
  \bibfield  {author} {\bibinfo {author} {\bibfnamefont {A.~L.}\ \bibnamefont
  {Fetter}},\ }\href@noop {} {\bibfield  {journal} {\bibinfo  {journal} {Phys.
  Rev. B}\ }\textbf {\bibinfo {volume} {22}},\ \bibinfo {pages} {1200}
  (\bibinfo {year} {1980})}\BibitemShut {NoStop}%
\bibitem [{\citenamefont {Baelus}\ \emph {et~al.}(2004)\citenamefont {Baelus},
  \citenamefont {Cabral},\ and\ \citenamefont {Peeters}}]{Baelus04}%
  \BibitemOpen
  \bibfield  {author} {\bibinfo {author} {\bibfnamefont {B.~J.}\ \bibnamefont
  {Baelus}}, \bibinfo {author} {\bibfnamefont {L.~R.~E.}\ \bibnamefont
  {Cabral}}, \ and\ \bibinfo {author} {\bibfnamefont {F.~M.}\ \bibnamefont
  {Peeters}},\ }\href@noop {} {\bibfield  {journal} {\bibinfo  {journal} {Phys.
  Rev. B}\ }\textbf {\bibinfo {volume} {69}},\ \bibinfo {pages} {064506}
  (\bibinfo {year} {2004})}\BibitemShut {NoStop}%
\bibitem [{\citenamefont {Kogan}(1994)}]{Kogan94}%
  \BibitemOpen
  \bibfield  {author} {\bibinfo {author} {\bibfnamefont {V.~G.}\ \bibnamefont
  {Kogan}},\ }\href@noop {} {\bibfield  {journal} {\bibinfo  {journal} {Phys.
  Rev. B}\ }\textbf {\bibinfo {volume} {49}},\ \bibinfo {pages} {15874}
  (\bibinfo {year} {1994})}\BibitemShut {NoStop}%
\bibitem [{\citenamefont {Kogan}(2007)}]{Kogan07}%
  \BibitemOpen
  \bibfield  {author} {\bibinfo {author} {\bibfnamefont {V.~G.}\ \bibnamefont
  {Kogan}},\ }\href@noop {} {\bibfield  {journal} {\bibinfo  {journal} {Phys.
  Rev. B}\ }\textbf {\bibinfo {volume} {75}},\ \bibinfo {pages} {064514}
  (\bibinfo {year} {2007})}\BibitemShut {NoStop}%
\bibitem [{\citenamefont {Stejic}\ \emph {et~al.}(1994)\citenamefont {Stejic},
  \citenamefont {Gurevich}, \citenamefont {Kadyrov}, \citenamefont {Christen},
  \citenamefont {Joynt},\ and\ \citenamefont {Larbalestier}}]{Stejic94}%
  \BibitemOpen
  \bibfield  {author} {\bibinfo {author} {\bibfnamefont {G.}~\bibnamefont
  {Stejic}}, \bibinfo {author} {\bibfnamefont {A.}~\bibnamefont {Gurevich}},
  \bibinfo {author} {\bibfnamefont {E.}~\bibnamefont {Kadyrov}}, \bibinfo
  {author} {\bibfnamefont {D.}~\bibnamefont {Christen}}, \bibinfo {author}
  {\bibfnamefont {R.}~\bibnamefont {Joynt}}, \ and\ \bibinfo {author}
  {\bibfnamefont {D.~C.}\ \bibnamefont {Larbalestier}},\ }\href@noop {}
  {\bibfield  {journal} {\bibinfo  {journal} {Phys. Rev. B}\ }\textbf {\bibinfo
  {volume} {49}},\ \bibinfo {pages} {1274} (\bibinfo {year}
  {1994})}\BibitemShut {NoStop}%
\bibitem [{\citenamefont {Bulaevskii}\ \emph {et~al.}(2011)\citenamefont
  {Bulaevskii}, \citenamefont {Graf}, \citenamefont {Batista},\ and\
  \citenamefont {Kogan}}]{Bulaevskii11}%
  \BibitemOpen
  \bibfield  {author} {\bibinfo {author} {\bibfnamefont {L.~N.}\ \bibnamefont
  {Bulaevskii}}, \bibinfo {author} {\bibfnamefont {M.~J.}\ \bibnamefont
  {Graf}}, \bibinfo {author} {\bibfnamefont {C.~D.}\ \bibnamefont {Batista}}, \
  and\ \bibinfo {author} {\bibfnamefont {V.~G.}\ \bibnamefont {Kogan}},\
  }\href@noop {} {\bibfield  {journal} {\bibinfo  {journal} {Phys. Rev. B}\
  }\textbf {\bibinfo {volume} {83}},\ \bibinfo {pages} {144526} (\bibinfo
  {year} {2011})}\BibitemShut {NoStop}%
\bibitem [{\citenamefont {H\"{a}nggi}\ \emph {et~al.}(1990)\citenamefont
  {H\"{a}nggi}, \citenamefont {Talkner},\ and\ \citenamefont
  {Borkovec}}]{Hanggi90}%
  \BibitemOpen
  \bibfield  {author} {\bibinfo {author} {\bibfnamefont {P.}~\bibnamefont
  {H\"{a}nggi}}, \bibinfo {author} {\bibfnamefont {P.}~\bibnamefont {Talkner}},
  \ and\ \bibinfo {author} {\bibfnamefont {M.}~\bibnamefont {Borkovec}},\
  }\href@noop {} {\bibfield  {journal} {\bibinfo  {journal} {Rev. Mod. Phys.}\
  }\textbf {\bibinfo {volume} {62}},\ \bibinfo {pages} {251} (\bibinfo {year}
  {1990})}\BibitemShut {NoStop}%
\bibitem [{\citenamefont {Bulaevskii}\ \emph {et~al.}(2012)\citenamefont
  {Bulaevskii}, \citenamefont {Graf},\ and\ \citenamefont
  {Kogan}}]{Bulaevskii11b}%
  \BibitemOpen
  \bibfield  {author} {\bibinfo {author} {\bibfnamefont {L.~N.}\ \bibnamefont
  {Bulaevskii}}, \bibinfo {author} {\bibfnamefont {M.~J.}\ \bibnamefont
  {Graf}}, \ and\ \bibinfo {author} {\bibfnamefont {V.~G.}\ \bibnamefont
  {Kogan}},\ }\href@noop {} {\bibfield  {journal} {\bibinfo  {journal} {Phys.
  Rev. B}\ }\textbf {\bibinfo {volume} {85}},\ \bibinfo {pages} {014505}
  (\bibinfo {year} {2012})}\BibitemShut {NoStop}%
\bibitem [{\citenamefont {Fulton}\ and\ \citenamefont
  {Dunkleberger}(1974)}]{Fulton74}%
  \BibitemOpen
  \bibfield  {author} {\bibinfo {author} {\bibfnamefont {T.~A.}\ \bibnamefont
  {Fulton}}\ and\ \bibinfo {author} {\bibfnamefont {L.~N.}\ \bibnamefont
  {Dunkleberger}},\ }\href@noop {} {\bibfield  {journal} {\bibinfo  {journal}
  {Phys. Rev. B}\ }\textbf {\bibinfo {volume} {9}},\ \bibinfo {pages} {4760}
  (\bibinfo {year} {1974})}\BibitemShut {NoStop}%
\bibitem [{\citenamefont {Smelyanskiy}\ \emph {et~al.}(1999)\citenamefont
  {Smelyanskiy}, \citenamefont {Dykman},\ and\ \citenamefont
  {Golding}}]{Smelyanskiy99}%
  \BibitemOpen
  \bibfield  {author} {\bibinfo {author} {\bibfnamefont {V.~N.}\ \bibnamefont
  {Smelyanskiy}}, \bibinfo {author} {\bibfnamefont {M.~I.}\ \bibnamefont
  {Dykman}}, \ and\ \bibinfo {author} {\bibfnamefont {B.}~\bibnamefont
  {Golding}},\ }\href@noop {} {\bibfield  {journal} {\bibinfo  {journal} {Phys.
  Rev. Lett.}\ }\textbf {\bibinfo {volume} {82}},\ \bibinfo {pages} {3193}
  (\bibinfo {year} {1999})}\BibitemShut {NoStop}%
\end{thebibliography}
%

\end{document}